\documentclass{article}
\usepackage{spconf,amsmath,graphicx}
\usepackage{multirow}
\usepackage{booktabs}
\usepackage{xcolor}
\usepackage{booktabs, makecell}
\usepackage{threeparttable}
\usepackage{amsfonts}
\usepackage{amssymb}
\usepackage{pifont}
\title{Multimodal Dyadic Impression Recognition via Listener Adaptive Cross-Domain Fusion}
\name{Yuanchao Li, Peter Bell, Catherine Lai}
\address{Centre for Speech Technology Research, University of Edinburgh, UK\\
        \{yuanchao.li, peter.bell, c.lai\}@ed.ac.uk}
\begin{document}
%
\maketitle
\begin{abstract}
As a sub-branch of affective computing, impression recognition, e.g., perception of speaker characteristics such as warmth or competence, is potentially a critical part of both human-human conversations and spoken dialogue systems. Most research has studied impressions only from the behaviors expressed by the speaker or the response from the listener, yet ignored their latent connection. In this paper, we perform impression recognition using a proposed listener adaptive cross-domain architecture, which consists of a listener adaptation function to model the causality between speaker and listener behaviors and a cross-domain fusion function to strengthen their connection. The experimental evaluation on the dyadic IMPRESSION dataset verified the efficacy of our method, producing concordance correlation coefficients of 78.8\% and 77.5\% in the competence and warmth dimensions, outperforming previous studies. The proposed method is expected to be generalized to similar dyadic interaction scenarios in affective computing.
\end{abstract}

\noindent\textbf{Index Terms}: dyadic interaction, impression recognition, listener adaptation, cross-domain fusion, multitask learning

\vspace{-3pt}
\section{Introduction}
\vspace{-3pt}
Besides planning spoken content, psychological research indicates that people also intentionally control their appearances and behaviors to leave different impressions on their interaction partners in scenarios such as selection, job performance, and leader–subordinate relationships. For example, people tend to express extraversion during job interviews and show agreeableness in dating \cite{swider2021first}. Social research has pointed out the importance of impressions in human interactions, where it is natural to perceive impressions from the partners via non-verbal behaviors such as eye gaze, body pose, speaking activity, and prosody variation \cite{biel2012youtube}. Impressions are often categorized in terms of the ``Big Five'' personality traits: \textit{extraversion, agreeableness, conscientiousness, neuroticism}, and \textit{openness}, or the formed opinions of social perception/cognition: \textit{warmth} and \textit{competence}. Warmth reflects intentions towards others and includes traits of being good-natured, trustworthy, tolerant, friendly, and sincere. Competence reflects the ability to enact intentions and means being capable, skillful, intelligent, and confident \cite{wang2021open}.

The procedure of impression recognition is similar to that of emotion recognition, consisting of two steps: feature extraction and classification/regression. Researchers have used many approaches to predict impressions from human behaviors, such as facial, vocal, and bodily expressions. \cite{aran2013one} used high-level features obtained from speaking activity, body and head motion, along with low-level features extracted from audio to predict the personality impressions. \cite{judge1999big} extracted visual and vocal features (e.g., eye gaze, head pose, speaking activity, and prosody variation) to characterize the social interaction. \cite{subramanian2016ascertain} used Electroencephalogram (EEG), Electrocardiogram (ECG), Galvanic Skin Response (GSR), and facial activity data to recognize personality. 

Unlike emotions that almost rely on the subjective feelings of the speaker, impressions also rely on how the listener perceives the speaker's expressions. However, most previous work recognized impressions by using only the speaker or listener behaviors, ignoring the fact that impressions are formed by both of the dyadic partners. Therefore, in this paper, we propose a dyadic impression recognition architecture using listener adaptive cross-domain fusion to capture and strengthen causal and related information from the speaker and listener as there is a perception gap that separates them into two domains of signal sources (explained in Sec 2).

\vspace{-3pt}
\section{Related Work}
\vspace{-3pt}
\noindent\textbf{Dyadic impression recognition.}
The majority of impression research focuses on personality impressions, yet more and more studies on other impression dimensions have emerged. The Noxi corpus \cite{cafaro2017noxi} was collected to investigate the relationship between observed non-verbal cues and the first impression formation of warmth and competence \cite{biancardi2017analyzing}. AMIGOS \cite{correa2018amigos} collected participants' multimodal behaviors expressed during watching videos for the assessment of various affective levels (e.g., valence, arousal, control, familiarity, liking). The Dyadic IMPRESSION dataset \cite{wang2021open} contains audio, visual, and physiological signals from both the speaker and listener for the recognition of warmth and competence. A critical issue, however, has long existed in this field yet not been solved: \textit{previous work either performed impression recognition from only speaker or listener behaviors, or failed to deal with their combination properly}. For example, 1) the studies on the SSPNet Speaker Personality Corpus \cite{mohammadi2012automatic} estimated speaker personalities using speaker audio and listener annotations but gave no consideration to listener responses, which were not collected; 2) AMIGOS provides only the participant (i.e., listener) signals without the videos (i.e., speaker), resulting in the situation that studies using this dataset for impression research can only obtain listener behaviors; 3) \cite{wang2021open} extracted both speaker and listener features for fusion but did not consider that the speaker signals (video sources) are identical for all listeners, bringing about redundancy in training data. Therefore, we aim to incorporate both speaker and listener features and fuse them properly.

\noindent\textbf{Feature fusion.}
Another major issue of impression recognition is feature fusion -- a general problem for affective computing tasks. Tensor fusion, which deals with features at the hidden-state level, is becoming dominant as it can better model synchrony, relatedness, and hierarchy of multimodal features. For example, attention-based and hierarchical tensor fusion methods have been investigated for features of different levels and proven useful in emotion recognition \cite{li2022fusing,tian2016recognizing}. However, we regard the fusion problem as more complex in impression recognition because \textit{impressions also depend on how listeners perceive speaker behaviors, which means there is a perception gap between the listener and speaker.} In particular, many studies on affective states are conducted in the scenarios of video watching or audio listening where there is no real-time interaction, making it difficult for the two parties (signal sources) to achieve consistency towards the target states \cite{tian2017recognizing}. Thus, in this work, we investigate effective fusion methods for such a cross-domain problem.

\vspace{-3pt}
\section{Data Preparation}
\vspace{3pt}
\noindent\textbf{Dyadic IMPRESSION dataset.}
This dataset contains multimodal signals from both the speaker and listener \cite{wang2021open}. The dataset consists of 31 dyads, in total 1,890 minutes of synchronized recordings of face videos, speech clips, eye gaze data, and peripheral nervous system physiological signals (e.g., EEG, ECG, blood volume pulse). 40 participants (listeners) watched 13 Noxi stimuli (speakers) and annotated their formed impressions in warmth and competence dimensions in real time. The labels are represented in a step-wise continuous manner, which means the participants were allowed to increase/decrease the label value once they felt an impression change. Thus, there is no value range limitation.

\noindent\textbf{Feature preparation.}
The provided features contain four modalities: audio, eye, face, and physio, from both the speaker and listener. Since the sample numbers of each modality are different, we then conducted resampling to make them the same as the label's. Finally, the 13 speakers have a total of 44,923 samples with 412 feature dimensions, while each of the 40 listeners has 44,923 samples with 68 feature dimensions, as in previous work \cite{wang2021open,li2022cross}.

\vspace{-3pt}
\section{Proposed Architecture}
\begin{figure*}[ht]
  \centering \includegraphics[width=0.895\linewidth]{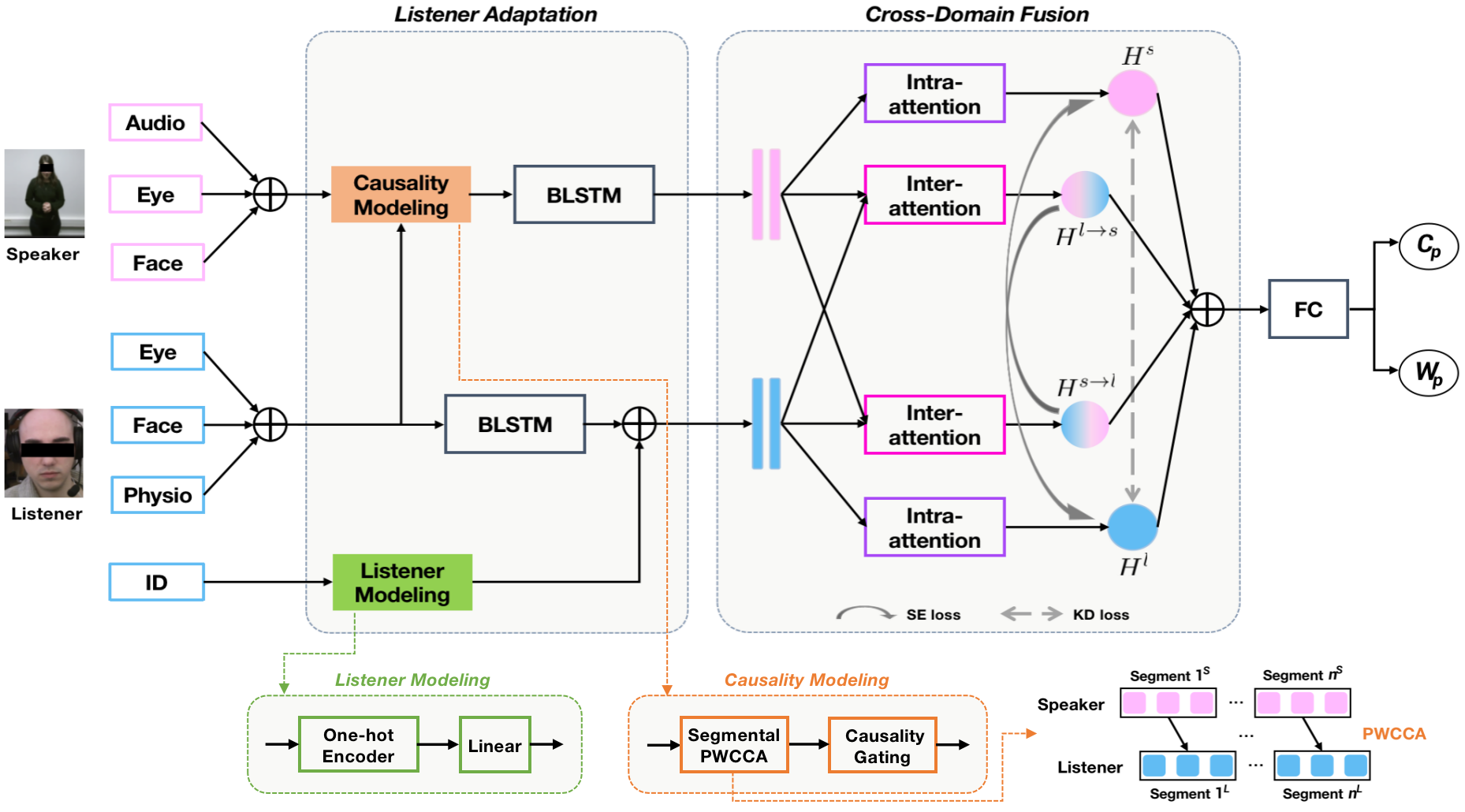}
  \caption{Proposed architecture using listener adaptive cross-domain fusion.}
  \label{fig:model}
\vspace{-12pt}
\end{figure*}

\vspace{-3pt}
To address the two issues stated in Sec 2, we propose a listener adaptive cross-domain architecture, shown in Fig~\ref{fig:model}. We first denote the concatenated features of the speaker and listener as $S$ and $L$, which are 412- and 68-dimensional feature sequences with the length of 44,923, respectively.

\vspace{-5pt}
\subsection{Listener Adaptation}
\noindent\textbf{Causality modeling.}
In dyadic impression recognition, the listener reacts according to the speaker's signals, e.g., with a faster heartbeat or a smile. This causal relationship is usually ignored in previous studies \cite{wang2021open}, leaving them unable to exploit the full potential of the features. To this end, we performed a novel causality modeling module consisting of segmental Projection-Weighted Canonical Correlation Analysis (PWCCA) and a causality gating process. \emph{Segmental PWCCA} first splits both speaker and listener features into $n$ segments (we used $n=$ 450, which means the length of the last segment is 23 and that of the previous 449 is 100), then calculates causality using every segment pair. This practice is inspired by the fact that in human interaction, listeners only respond to the speaker behaviors that they have just received in a short period of time \cite{knapp2013nonverbal}. For the causality calculation, we used PWCCA \cite{morcos2018insights}, a variant of CCA, which gives higher weights for directions accounting for a higher proportion of the input. This is useful for finding the salient parts in the speaker segments that have high correlations with the listener segments, which we regard as high-level causality. Next, \emph{causality gating} assigns the calculated PWCCA causality as the weight to each speaker segment to form causality-weighted speaker features as $S^W=\{s_{1}^{w_{1}}, s_{2}^{w_{2}}, \cdots, s_{n}^{w_{n}}\}$, where $s_{i}^{w_{i}}= w_{i}s_{i}$ ($s_{i}$: the $i$th speaker segment; $w_{i}$: the $i$th PWCCA causality).

\noindent\textbf{Listener modeling.}
Considering the annotation bias introduced when assessing the speaker recordings by different listeners, we propose a listener modeling module to incorporate listener IDs. The IDs were first transformed into one-hot embeddings by the one-hot encoder, followed by a linear layer to encode the listener information and reshape the embeddings to the same size as other listener features for concatenation.

After causality modeling, we fed the speaker features to a Bi-directional Long Short-Term Memory (BLSTM) network. Meanwhile, the listener features were fed to another BLSTM network (except for the listener ID, as it has no temporal information) and then concatenated with the listener ID.

\vspace{-5pt}
\subsection{Cross-Domain Fusion}
\noindent\textbf{Cross-domain attention.}
Following the listener adaptation function, a structured cross-domain attention consisting of \textbf{intra-attention} and \textbf{inter-attention} networks, aggregated information from the BLSTM hidden states and produced four fixed-length encodings: $H^{s}$, $H^{l}$, $H^{s\rightarrow{l}}$, and $H^{l\rightarrow{s}}$. For both intra- and inter-attention, we used multi-head self-attention:
\begin{align}
    H &= MultiHead(Q, K, V)W^O \\
        &= Concat(head_1, ..., head_m) \\
    head_i &= Attention(QW_i^Q,KW_i^K,VW_i^V) 
\end{align}
where $W^O$, $W_i^Q$, $W_i^K$, and $W_i^V$ are trainable parameters. $Q$, $K$, and $V$ represent the query, key, and value, respectively. For inter-attention, $Q$ is from one domain (the speaker or listener), while $K$ and $V$ are from the other. For intra-attention, the three parameters are from the same domain. The value of $m$ is 16, and $H$ denotes the concatenated multi-head representations: $H^{s}$, $H^{l}$, $H^{s\rightarrow{l}}$, and $H^{l\rightarrow{s}}$. Next, we concatenated the four representations and passed them to a Fully-Connected (FC) network with a ReLU activation function to generate the recognition outputs $C_p$ and $W_p$, which are competence and warmth predictions.

We used cross-domain attention because the speaker and listener signals can be regarded as different domains in this dataset. The impression labels have high correlations with listener features but low correlations with the speaker's \cite{wang2021open}, indicating that their relatedness is not obvious. This phenomenon is plausible as the listener responds to the speaker recordings without real interaction. Thus, we need inter-attention to find relevant features between the two domains.

Inter-attention has been adopted in affective computing over recent years \cite{li2022fusing}. It exchanges key-value pairs in multi-head self-attention. As shown in Fig~\ref{fig:model}, $H^{s\rightarrow{l}}$ denotes speaker-attended listener features and $H^{l\rightarrow{s}}$ is the reverse. However, there may be other useful information from individual modalities ignored by the inter-attention. We used intra-attention to resolve this issue. Intra-attention focuses on salient information in each respective signal domain towards impression recognition and generates $H^{s}$ and $H^{l}$ as the hidden representations. The four representations were then concatenated for the final non-linear combination.

\noindent\textbf{Cross-domain regularization.}
To reduce the discrepancy and further regulate the relatedness between the two different domains, we designed a cross-domain regularization that has a \textbf{Knowledge Distillation (KD)} loss $\mathcal{L}_{kd}$ and a \textbf{Similarity Enhancement (SE)} loss $\mathcal{L}_{se}$. Knowledge distillation is a deep learning technique used for training a small network under the guidance of a trained network \cite{romero2014fitnets}. Though this technique is widely used in model training, recent work has been inspired to apply it to transfer knowledge between hidden representations \cite{Huang2021AudioOrientedMM}. Considering the fact that the multimodal signals from the speaker and listener have weak relatedness, we used a KD loss to transfer the information from the other domain. Unlike inter-attention, which directly attends one domain to the other, which we call ``hard relatedness'', the KD loss enables indirectly learning multimodal knowledge with minor changes in a ``soft'' way: $H^{l}$ and $H^{s}$ can absorb information from each other to some extent while still maintaining their independence. We calculated the Mean Squared Error (MSE) for the KD loss (note that the two MSE calculations are identical):
\begin{align}
    \mathcal{L}_{kd} &= MSE(H^{s}, H^{l}) + MSE(H^{l}, H^{s})
\end{align}
To ensure the inter-attention representations have successfully learned the information from the other domain, we applied an SE loss. For example, minimizing the distance between $H^{l\rightarrow{s}}$ and $H^{l}$ to align the two representations means that the listener information has been attended to the speaker's by inter-attention. We used Kullback–Leibler (KL) divergence for this purpose (note that the variables have been converted to probability distributions using softmax):
\begin{align}
    \mathcal{L}_{se} &= KL(H^{s\rightarrow{l}}, H^{s}) + KL(H^{l\rightarrow{s}}, H^{l})
\end{align}
The reasons why we chose MSE and KL divergence are: 1) MSE generally outperforms KL divergence in knowledge distillation \cite{kim2021comparing}; 2) KL divergence is good at calculating the distance between two distributions in the same probability space and is popular for similarity measurement \cite{goldberger2003efficient}, so we expect it to enhance the similarity in the cross-domain situation. We also tried exchanging MSE and KL divergence for KD and SE but found a small decrease in the warmth dimension.

Finally, the concatenated representations were fed to an FC network, which contains a linear layer with 16 neurons, followed by a ReLU activation function, a dropout layer with 0.5 probability, and two parallel linear layers to generate predictions using multitask learning. The prediction task was optimized by the following objective function:
\begin{align}
    \mathcal{L}_{pred} &= MSE(C_{p}, C_{g}) + MSE(W_{p}, W_{g})
\end{align}
where $C_{p}$ and $W_{p}$ are the predictions of competence and warmth, and $C_{g}$ and $W_{g}$ are the corresponding groundtruth.

\vspace{-3pt}
\section{Experimental Evaluation}
\vspace{-3pt}
\subsection{Implementation}
The model was built using Pytorch and optimized using the Adam method. The learning rate was set at 1e-3 and reduced by half every 20 epochs. The model was trained for 40 epochs by minimizing the overall loss:
\begin{align}
    \mathcal{L} &= \mathcal{L}_{pred} + \mathcal{L}_{kd} + \mathcal{L}_{se}
\end{align}
We randomly used 80\% of the data for training, 10\% for validation, and 10\% for testing. The performance was evaluated using the Concordance Correlation Coefficient (CCC).

\vspace{-5pt}
\subsection{Results and Discussion}
\vspace{-5pt}
We compare our results with the only two studies on this dataset and present an ablation study where we removed each component. Table~\ref{tab:result} shows that: 1) our proposed method achieves the best results by largely increasing the CCC; 2) the removal of each of the components causes a decrease in performance, which in turn proves their effectiveness, and causality modeling contributes the most as the decrease is the largest; 3) inter-attention works better than intra-attention, suggesting that the cross-domain relevancy is learned and contributes to the recognition. To verify if LA really modeled the latent relationship, we compare the contribution of cross-domain fusion (i.e., performance decrease) with and without LA in Table~\ref{tab:LA}. It shows that cross-domain fusion contributes more with LA, indicating that with the help of LA modeling the latent causality and listener information, it becomes easier for cross-domain fusion to capture the speaker-listener relatedness. In particular, the contribution of SE loss changes the most, indicating the relatedness between $H^{s}$ and $H^{l}$ encoded from two respective domains has been modeled by LA.

\begin{figure}[ht]
\vspace{-5pt}
  \centering
  \includegraphics[width=\linewidth]{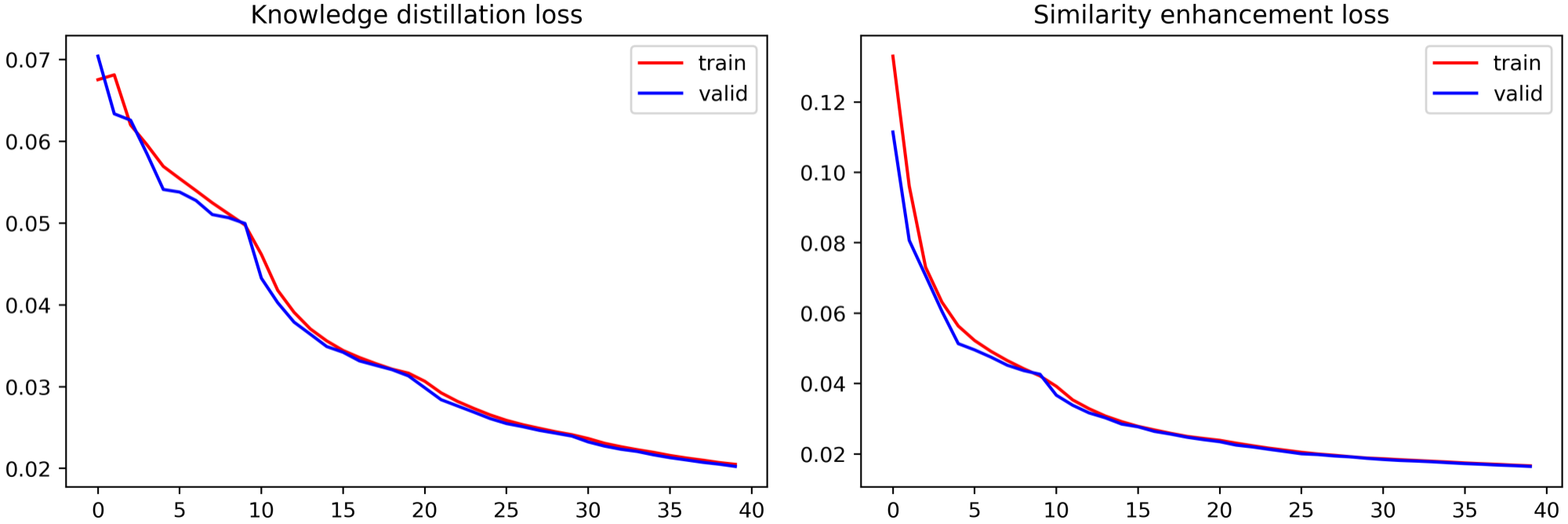}
  \caption{Trends of regularization losses.}
  \label{fig:trend}
\vspace{-7pt}
\end{figure}

Moreover, in Fig~\ref{fig:trend}, we observe that both KD and SE losses show a decreasing trend. However, we saw an increasing trend when removed from back-propagation, which also demonstrates their usefulness (figure omitted for brevity).

\begin{table}[t]
\vspace{-7pt}
\centering
  \caption{Results. (-): removal of the component.}
  \label{tab:result}
  \begin{tabular}{lcc}
    \toprule
    \textbf{Method} & \textbf{Competence} & \textbf{Warmth} \\
    \midrule
    Wang et al. \cite{wang2021open} & 73.7 & 75.1 \\
    Li et al. \cite{li2022cross} & 77.0 & 74.8 \\
    \textit{Ours} & \textbf{78.8} & \textbf{77.5} \\
    \midrule
    (-) Causality modeling & 77.3 \color{cyan}$\downarrow$1.5 & 75.6 \color{cyan}$\downarrow$1.9\\
    (-) Listener modeling & 77.9 \color{cyan}$\downarrow$0.9 & 76.8 \color{cyan}$\downarrow$0.7\\
    (-) Inter-attention & 77.7 {\color{cyan}$\downarrow$1.1} & 76.4 {\color{cyan}$\downarrow$1.1}\\
    (-) Intra-attention & 77.8 \color{cyan}$\downarrow$1.0 & 76.6 \color{cyan}$\downarrow$0.9\\
    (-) KD loss & 77.9 \color{cyan}$\downarrow$0.9 & 76.9 \color{cyan}$\downarrow$0.6\\
    (-) SE loss & 77.7 \color{cyan}$\downarrow$1.1 & 76.8 \color{cyan}$\downarrow$0.7\\
    \bottomrule
\end{tabular}
\vspace{-5pt}
\end{table}

\begin{table}[t]
\vspace{-5pt}
\centering
\caption{Performance decrease of removing cross-domain fusion components w/ and w/o Listener Adaptation (LA).}
\label{tab:LA}
\begin{tabular}{lcccccccc}
\toprule
\multicolumn{1}{l|}{\multirow{2}{*}{\textbf{LA}}} & \multicolumn{2}{c}{Inter-attn} & \multicolumn{2}{c}{Intra-attn} & \multicolumn{2}{c}{KD loss} & \multicolumn{2}{c}{SE loss} \\ \cline{2-9} 
\multicolumn{1}{c|}{} & C & W & C & W & C & W & C & W \\ \hline
\multicolumn{1}{c|}{w/} & \color{cyan}1.1 & \color{cyan}1.1 & \color{cyan}1.0 & \color{cyan}0.9 & \color{cyan}0.9 & \color{cyan}0.6 & \color{cyan}1.1 & \color{cyan}0.7 \\
\multicolumn{1}{c|}{w/o} & 0.9 & 0.7 & 0.7 & 0.8 & 0.7 & 0.5 & 0.4 & 0.3 \\
\bottomrule
\end{tabular}
\vspace{-10pt}
\end{table}

\vspace{-5pt}
\section{Conclusions}
\vspace{-5pt}
In this paper, we propose a listener adaptive cross-domain architecture to address long-existing problems in dyadic impression recognition. This architecture consists of a listener adaptation function and a cross-domain fusion function to model the causality between speaker and listener behaviors and capture their relatedness. The experimental evaluation shows that both functions help and the fusion works better with the listener adaptation. We expect the proposed architecture can be generalized to similar dyadic interaction scenarios and will test it on other datasets in our future work.

\bibliographystyle{IEEEbib}
\bibliography{IEEE}

\end{document}